\documentclass[aps]{revtex4}
\usepackage{graphicx}
\usepackage{dcolumn}

\newcommand{\ep}{\epsilon}

\newcommand{\lraw}{\longrightarrow}

\newcommand{\pa}{\partial}
\newcommand{\td}{\tilde}

\newcommand{\beq}[1]{\begin{eqnarray}\label{#1}}
\newcommand{\eeq}{\end{eqnarray}}
\newcommand{\net}{{\cal N}=1}
\newcommand{\gym}{g_{_{\rm YM}}}

\newcounter{saveeqn}
\newcommand{\alphaeqn}{\setcounter{saveeqn}{\value{equation}}}
\stepcounter{saveeqn} 
\renewcommand{\theequation}{\mbox{\arabic{saveeqn}-\alpha{equation}}}
\newcommand{\reseteqn}{\setcounter{equation}{\value{saveeqn}}}
\renewcommand{\theequation}{\arabic{equation}}

\begin{document}

\title{Confinement of ${\cal N}=1$ Super Yang-Mills from Supergravity}

\author{Xiao-Jun Wang}

\email{wangxj@itp.ac.cn} \affiliation{Interdisciplinary center for theoretical
study, \\ University of Science and Technology of China, \\ Anhui, Hefei,
230026, P.R.China}

\author{Sen Hu}
\email{shu@ustc.edu.cn} \affiliation{Interdisciplinary center for theoretical
study \\ and Department of Mathematics, University of Science and
Technology of China, \\
Anhui, Hefei, 230026, P.R. China}

\begin{abstract}
We calculate circular Wilson loop of pure ${\cal N}=1$ super Yang-Mills
from Klebanov-Strassker-Tseytlin solution of supergravity and proposed
gauge/gravity duality. The calculation is performed numerically via
searching minimal surface of string worldsheet. It is shown that Wilson
loop exhibits area law for large radius such that ${\cal N}=1$ super
Yang-Mills is confined at large distance. Meanwhile, Wilson loop exhibits
logarithmic behavior for small radius and indicates asymptotical freedom
of ${\cal N}=1$ super Yang-Mills.
\end{abstract}

\pacs{11.15.-q,4.50.+h,4.65.+e,11.27.+d}

\maketitle

\section{Introduction}

Physicists now believe non-abelian gauge theory exhibits some
dramatic behaviors at long distance, such as confinement and mass
gap, etc.. It is difficult to study these properties using
traditional quantum field theory (QFT) method due to its
nonpertubative characters. However, the remarkable success of
AdS/CFT correspondence reveals a practical approach to study
strong coupling gauge theory\cite{Mald98,GKP98,Witten98}. The
similar correspondence is believed to exist in broad class of
theories even with less supersymmetry and non-conformal theories
(so-called gauge/gravity duality). According to continuously
breaking supersymmetry, we can expect to study strong-coupling
properties of non-abelian gauge theory via dual weak-coupling
gravity. So far, several nonsingular supergravity (SUGRA) solution
have been constructed. They preserve 1/2 or 1/4 of the maximal
supersymmetry and are conjectured dual to $d=4$, ${\cal
N}=2$\cite{GKM01} or ${\cal N}=1$\cite{MN01,KT00,KS00} super
Yang-Mills (SYM) theories. In particular, the case ${\cal N}=1$
super SYM is very interesting, because it possesses the common
features of non-abelian gauge theory at long distance but is hard
to be study using usual QFT method. For this SYM, two dual SUGRA
solution were found:
\begin{itemize}
\item The Maldacena-Nu$\td{\rm n}$ez (MN) solution\cite{MN01}:
It corresponds to a large ($N$) number of D5-branes wrapped on a
supersymmetric two-cycle inside a Calabi-Yau threefold. Meanwhile, 6d SYM
theory onto D5-branes is reduced to pure $d=4,\;\net$ $SU(N)$ SYM in the
IR.
\item The Klebanov-Strassler-Tseytlin (KST)
solution~\cite{KT00,KS00}: It describes the geometry of the warped
deformed conifold when one places M D3-branes and N fractional
D3-branes at the apex of the conifold. It is dual to a certain
$\net$ supersymmetric $SU(N+M)\times SU(M)$ gauge theory. If $M$
is a multiple of $N$, then this theory flows to $SU(N)$ in the IR,
via a chain of duality cascade which reduces the rank of the gauge
group by $N$ units at each cascade jump. Thus at the end of the
duality cascade the gauge theory is effectively pure $\net$
$SU(N)$ SYM.
\end{itemize}
The purpose of this paper is to study confinement of $\net$ SYM
from KST solution. We will give a brief comments on MN solution
at the end of the paper.

The natural criterion for confinement of pure Yang-Mills theory
is expectation value of large Wilson loop satisfying area
law\cite{WTM}. The evaluation on Wilson loop of ${\cal N}=4$ SYM
from supergravity in $AdS$ space has been studied in various
aspects. It was proposed in\cite{Mald98b,RY98} that Wilson loops
of the CFT can be described in AdS by
 \beq{1}
   <W({\cal C})>\;\sim\;\lim_{\Phi\to\infty}e^{-(S_{\Phi}-l\Phi)},
 \eeq
where $S$ is the proper area of a fundamental string world-sheet
which lies on the loop ${\cal C}$ on the boundary of AdS, $l$ is
the total length of the Wilson loop and $\Phi$ is the mass of the
$W$ boson. The presence of the term $l\Phi$ is to subtract
divergence from infinity mass of the $W$ boson. It corresponds to
UV divergence in QFT. The evaluation for the rectangular Wilson
loop was performed in refs.\cite{Mald98,RY98} and for simpler
circular loop was performed in refs.\cite{BCFM99,DGO99}. There
are no any surprise results due to conformal invariance. The more
non-trivial results is from calculation of operator product
expansion for the Wilson loop when probed from a distance much
larger than the size of the loop\cite{BCFM99}, and from
calculation on Wilson loop correlator\cite{GOphase}. The
extension of Eq.~(\ref{1}) to KST background is in principle
direct. However, since KST background is much more complicated
than AdS space, we can not expect to obtain any analytic results.
So that although there is a simple scaling argument that large
Wilson loops of $\net$ SYM resulted by KST solution satisfies the
area law\cite{HKO01}, the direct verification is still lacked so
far. In this paper we use numerical tools to search minimal
surface of string world-sheet in KST background which lies on the
Wilson loop on the boundary of KST space. In order to simplify
the calculation, we consider circular Wilson loop of $\net$ SYM in
Euclidean space. The proper area of string world-sheet can be
expressed as a function of radius of circular loop. We will
evaluate both of cases for small and large radius. It can be
expected that behavior of Wilson loop exhibits the following
conclusions,
\begin{itemize}
\item The confinement of $\net$ SYM at large distance.
\item The asymptotical freedom of $\net$ SYM at short distance.
\item Phase transition from short distance to large distance.
\end{itemize}
This study can also be treated as an examination for duality
between SUGRA KST solution and $\net$ SYM.

The paper is organized as follows. In Sec.\ref{sec2}, we extend
to calculation on Wilson loop in AdS space to KST background and
derive Euler-Lagrangian equation which governs minimal surface of
string world-sheet. In Sec.\ref{sec3}, We show some numerical
results such as curve of Wilson loop vs. radius of the loop. The
physical results are obtained via those numerical results. A
brief summary will be devoted in Sec.\ref{sec4} and several
details on numerical evaluation are presented in the Appendix.

\section{Wilson Loop of $\net$ SYM from KST solution}
\label{sec2}

The Wilson loop operator in $\net$ SYM is
 \beq{2}
  W({\cal C})=\frac{1}{N}Tr P e^{i\oint_{\cal C} A},
 \eeq
where ${\cal C}$ denotes a closed loop in spacetime, and the
trace is taken over the fundamental representation of the gauge
group. The expectation of the Wilson loop can be calculated by
supergravity in terms of Eq.~(\ref{1}) directly, with action $S$
is given by
 \beq{3} S=\frac{1}{2\pi\alpha'}\int d^2\sigma\sqrt{-g}, \eeq
where $\sigma^1$ and $\sigma^2$ parameterize fundamental string
(Euclidean) world-sheet,
 \beq{4} g_{\alpha\beta}=G_{MN}\pa_\alpha X^M\pa_\beta X^N, \eeq
is induced metric on the world-sheet. In this paper, we focus our
attention on nonsingular KST background such that $G_{MN}$ is
given by
 \beq{5}
 ds_{10}^2&=&h^{-1/2}(\tau)dx_ndx_n+h^{1/2}(\tau)ds_6^2,
    \nonumber\\
 h(\tau)&=&\alpha\int_\tau^\infty dx
     \frac{x\coth{x}-1}{\sinh^2{x}}(\sinh{(2x)}-2x)^{1/3}
    \nonumber \\
    &=&\alpha I(\tau),
 \eeq
where $\tau$ is the radial parameter of the transverse space,
$\alpha=2^{2/3}(g_sN\alpha')^2\ep^{-8/3}$. $ds_6^2$ is the metric
of the deformed conifold\cite{DC},
 \beq{6}
  ds_6^2=\frac{1}{2}\ep^{4/3}K(\tau)
  \left[\frac{1}{3K^3(\tau)}(d\tau^2+(g^5)^2)
     +\cosh^2{(\frac{\tau}{2})}\sum_{i=3}^4(g^i)^2
   +\sinh^2{(\frac{\tau}{2})}\sum_{i=1}^2(g^i)^2
 \right] \eeq
where $\ep$ parameterizes length of the deformed conifold and
possesses mass dimension $-3/2$, $K(\tau)$ is defined by
 \beq{7}
 K(\tau)=\frac{(\sinh{(2\tau)}-2\tau)^{1/3}}{2^{1/3}\sinh{\tau}},
 \eeq
and the 1-form basis $g^i$ parameterize transverse five-dimension space.
To simplify complicated calculation we do the following notions:
\begin{itemize}
\item Set angle parameters of transverse space as constants
such that $g^i\equiv 0$. The Wilson loop in ${\cal N}=4$ SYM with
non-constant angle parameters has been considered in ref.\cite{TZ02}. When
we extend this consideration to $\net$ SYM calculation becomes very
difficult.
\item To define a new dimensionless radial parameter
$r=e^{-\tau/3}$, $0\leq r\leq 1$. The boundary of KST space is
$r=0$.
\item We consider that $\net$ SYM is defined in Euclidean space,
and circular Wilson loop is located at $x^1-x^2$ plane. Since we are
aiming at finding a minimal surface, the string world-sheet can be
considered as a rotational surface. Then we define
 \beq{8} x^1&=&\sqrt{\frac{3}{2}\alpha}\ep^{2/3}f(r)\cos{\theta},
   \nonumber \\
      x^2&=&\sqrt{\frac{3}{2}\alpha}\ep^{2/3}f(r)\sin{\theta},
 \eeq
with a dimensionless function $f(r)$. Consequently the string
world-sheet is parameterized by $r$ and $\theta$,
 \beq{9}
   ds^2=\frac{3}{2}\sqrt{\alpha}\ep^{4/3}I^{-1/2}(r)\left\{
    [f'^2(r)+\frac{1}{r^2}I(r)K^{-2}(r)]dr^2+f^2(r)d\theta^2\right\},
 \eeq
where prime denotes derivative on $r$.
\end{itemize}

According to the above notions, the action defined in
Eq.~(\ref{3}) is rewritten to
 \beq{10}
  S=\frac{3}{2^{2/3}}g_sN\int dr\cdot I^{-1/2}
   f\sqrt{f'^2+G(r)},
 \eeq
with
 \beq{11} G(r)=\frac{1}{r^2}I(r)K^{-2}(r). \eeq
From action~(\ref{10}) we obtain the Euler-Lagrangian equation
which govern minimal surface of string world-sheet,
 \beq{12}
   f''-\frac{1}{2}(\ln{IG})'f'-\frac{I'}{2IG}f'^3-
   \frac{1}{f}(f'^2+G)=0.
 \eeq
The minimal surface implies the initial conditions of
Eq.~(\ref{12}) should be
 \beq{13} f(0)=R,\hspace{0.5in}f'(0)=0. \eeq
In order to show that this initial condition is taken care of properly,
let us find asymptotical solution of Eq.~(\ref{12}) for $r\to 0$. Noticing
the expressions,
 \beq{14}\left.\begin{array}{ccc}
   \displaystyle I(r)&\stackrel{r\to 0}{\lraw}&
  -\frac{3}{16\cdot 2^{1/3}} r^4(1+12\ln{r})+O(r^{10}) \\
   & & \\
   K(r)&\stackrel{r\to 0}{\lraw}& 2^{1/3}r+O(r^7)
   \end{array}\right\}
 \quad \Rightarrow \quad G(r)\stackrel{r\to 0}{\lraw}
   -\frac{3}{32}(1+12\ln{r}), \eeq
at limit $r\to 0$, the Euler-Lagrangian equation~(\ref{12})
approaches to
 \beq{15} ff''-(\frac{2}{r}+\frac{1}{r\ln{r}})ff'
   +\frac{9}{8}\ln{r}+\frac{3}{32}=0. \eeq
Considering the initial condition~(\ref{13}), we obtain
asymptotical solution of the above equation as follows
 \beq{16} f(r\to 0)&=&R+\frac{9}{16R}r^2\ln{r}
  -\frac{15}{64R}r^2+O(r^3), \nonumber \\
  f'(r\to 0)&=&\frac{9}{8R}r\ln{r}+\frac{3}{32R}r+O(r^2).
 \eeq

In addition, the minimal surface also implies that $f(r)$ must vanish at a
point $r=r_0$ with $0<r\leq r_0< 1$, but $f'(r)$ is divergent at this
point. So that the numerical evaluation will not be valid near this point,
and we have to find asymptotical solution analytically for $r\to r_0$,
 \beq{17}
   f(r\to r_0)=c(r_0-r)^{1/2}+O((r_0-r)^{3/2}), \eeq
with constant $c^2=8I(r_0)G(r_0)/I'(r_0)$.

Another aspect which need to be dealt with manually is
renormalization of divergence. It is caused by integration in
Eq.~(\ref{10}) when $r\to 0$. This divergence should be
subtracted from numerical result on proper area of string
world-sheet. Explicitly, using Eqs.~(\ref{14}) (\ref{16}) and
taking a cut-off $\lambda\to 0$, we obtain the divergence as
follows
 \beq{18} \int dr\cdot I^{-1/2}f\sqrt{f'^2+G(r)}=
 \frac{R}{2^{1/3}\lambda}+\frac{9}{2^{1/3}\cdot 32R}\lambda+O(\lambda^2)
 +{\rm other\;\;terms\;\;independent\;\;of}\;\;\lambda.
 \eeq
Here we include sub-sub leading term which is proportional to
$\lambda/R$ such that it will strictly vanish at limit $\lambda\to
0$. In numerical evaluation, however, the cut-off $\lambda$ is no
longer very small. Then this term will play a role for small $R$.

According to the above discussions, the Euler-Lagrangian
equation~(\ref{12}) can be solved numerically. The consistence of
numerical calculation will be checked via match initial
conditions (\ref{13}) and (\ref{17}). Consequently, the proper
area~(\ref{10}) of string world can be obtained after subtracting
$\lambda$-dependent terms in Eq.~(\ref{18}). Since difference
equation associating equation~(\ref{12}) is ill-defined for $s\to
s_0$, some tricks on numerical evaluation are needed. We put all
details on numerical evaluation in appendix of this paper.

\section{Numerical results and discussions}
\label{sec3}

We first show some numerical results:
\begin{itemize}
\item Fig.~1 includes some curves of $f(r)$ vs. $r$, which
represents some solutions of Euler-Lagrangian equation~(\ref{12})
with different initial values $R$.
\begin{figure}[hptb]
\parbox{3.5in}{
\label{f1}
   \centering
   \includegraphics[width=3in]{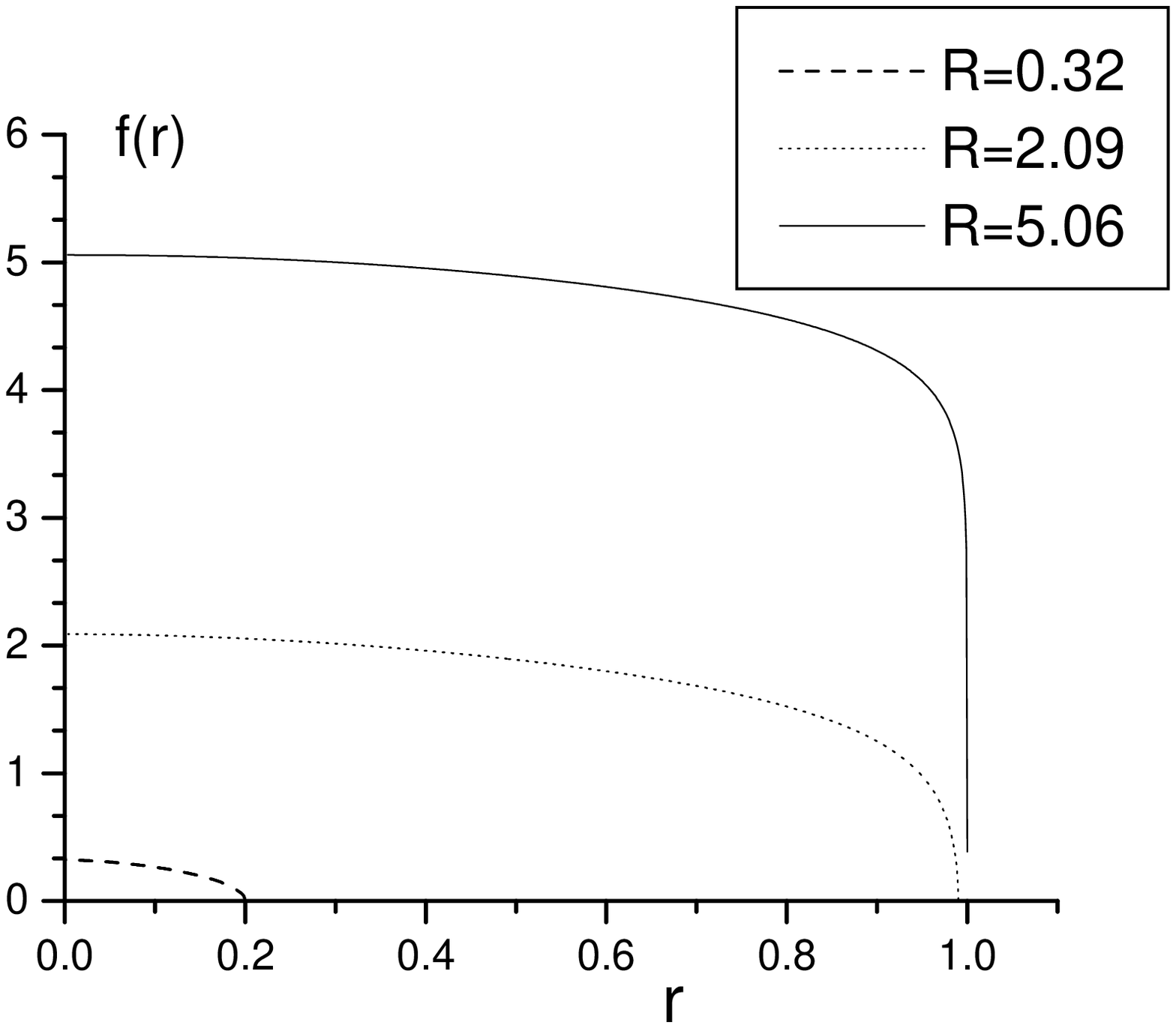}
\begin{minipage}{3in}
   \caption{Some solutions of Euler-Lagrangian equation~(\ref{12})
   with different initial values. The dash line denotes $R=0.32$,
   the dot line denotes $R=2.09$ and solid line denotes $R=5.06$.}
\end{minipage}}
\parbox{3.5in}{\label{f2}
   \centering
   \includegraphics[width=3in]{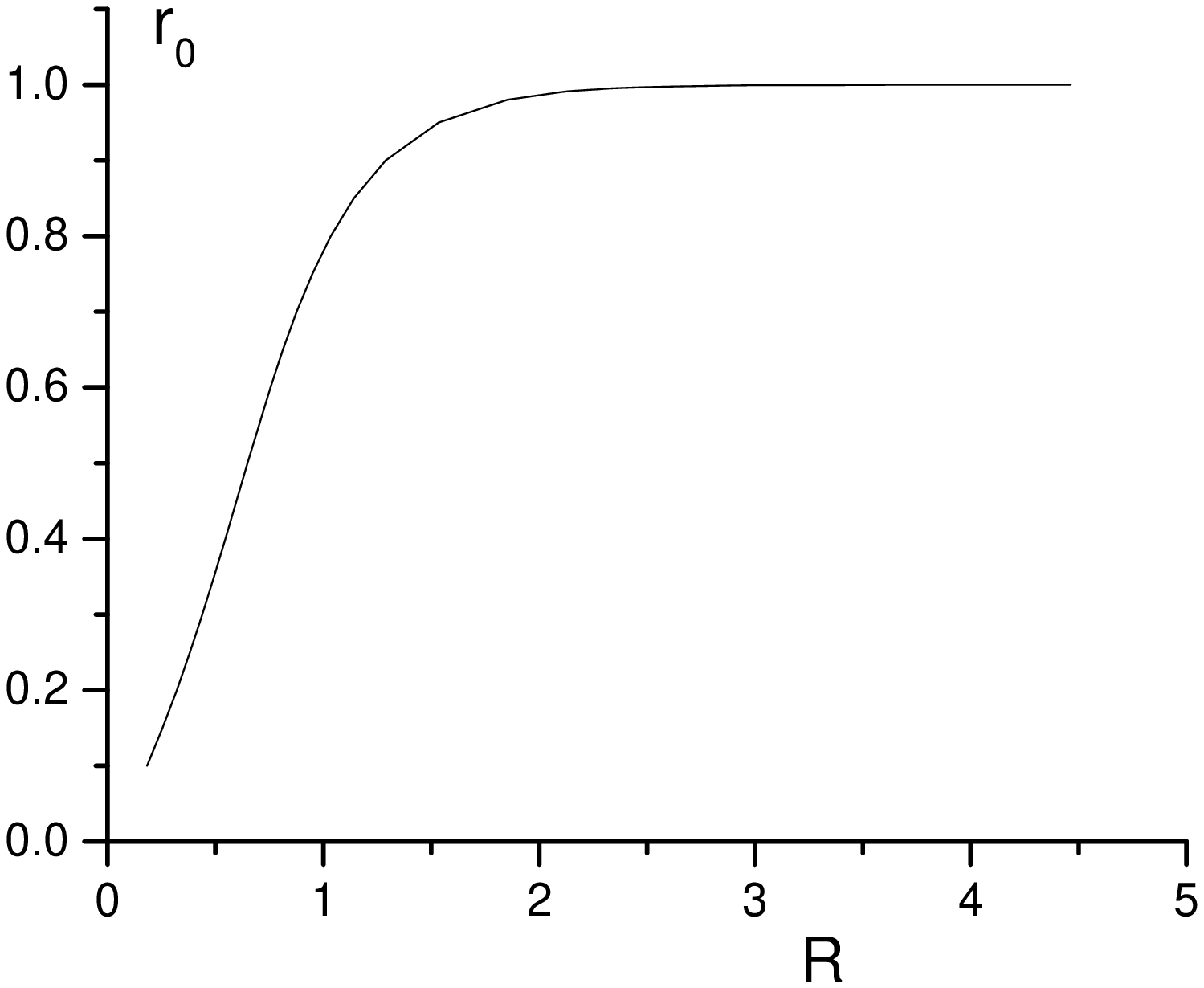}
\begin{minipage}{3in}
   \caption{Curve of $r_0$ vs. $R$, where $r_0$ is
zero-point of $f(r)$ and $R$ is initial value.}
\end{minipage}}
\end{figure}
\item Fig.~2 denotes curve of $r_0$ vs. $R$, where $r_0$ is
zero-point of $f(r)$ and govern integral region in
Eq.~(\ref{10}), i.e., $0\leq r\leq r_0$.
\item Fig.~3 and fig.~4 show that the resulted numerical data
and related fit curves on proper area $A$ of string world-sheet
vs. radius $R$ of Wilson loop at large distance and short distance
respectively. The unit of area $A$ is taken over $3g_sN/2^{2/3}$.
\begin{figure}[hptb]
\parbox{3.5in}{
\label{f3}
   \centering
   \includegraphics[width=3in]{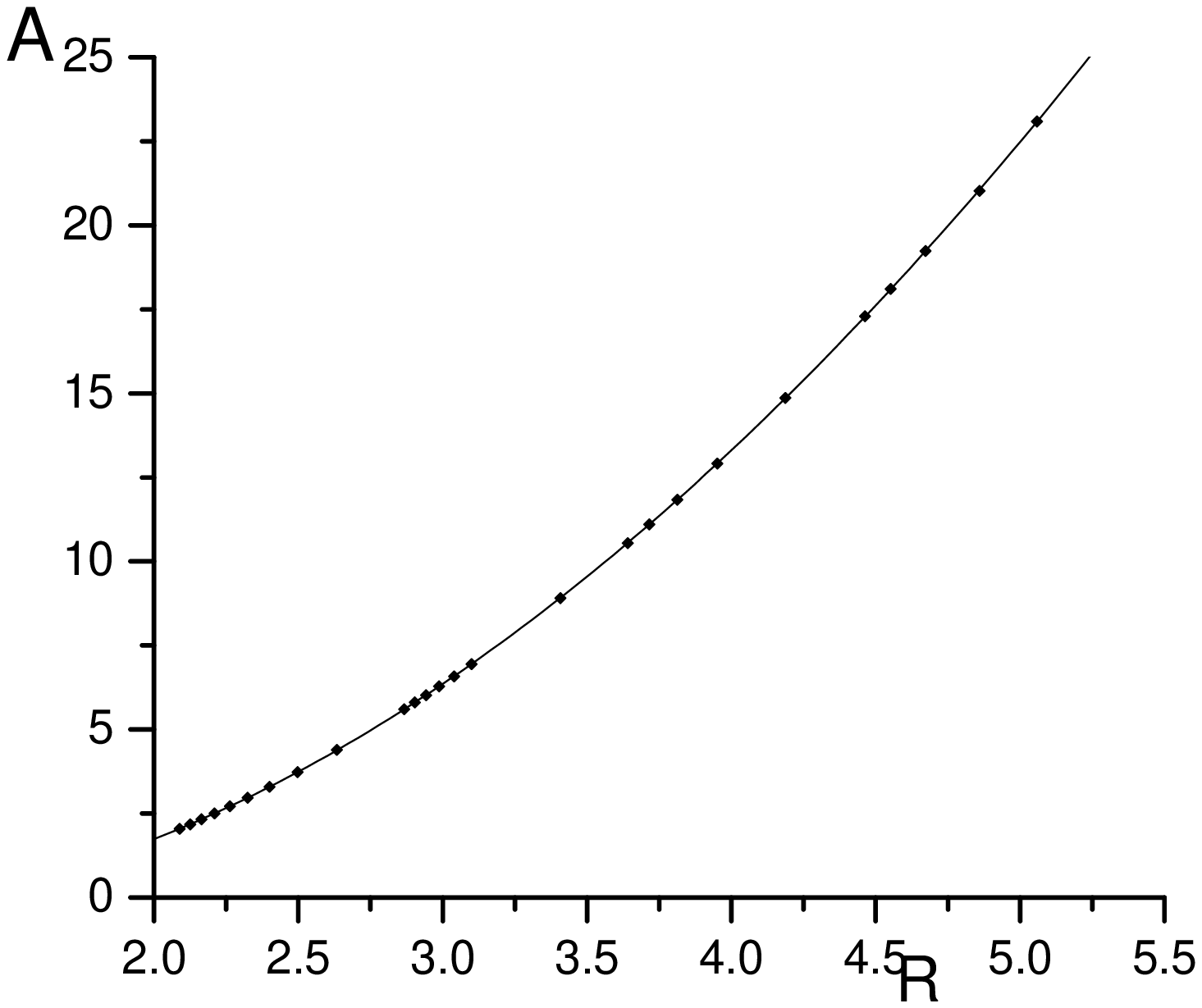}
\begin{minipage}{3in}
   \caption{Curves of proper area $A$ (in unit $3g_sN/2^{2/3}$) of
   string world-sheet vs. radius $R$ of Wilson loop at larger
   distance. The diamonds denote numerical results, and solid line is
   the fit function, $A=-1.5126+0.2941R+1.04R^2-2.162\ln{R}$
   with $\chi^2=0.001277$.}
\end{minipage}}
\parbox{3.5in}{
\label{f4}
   \centering
   \includegraphics[width=3in]{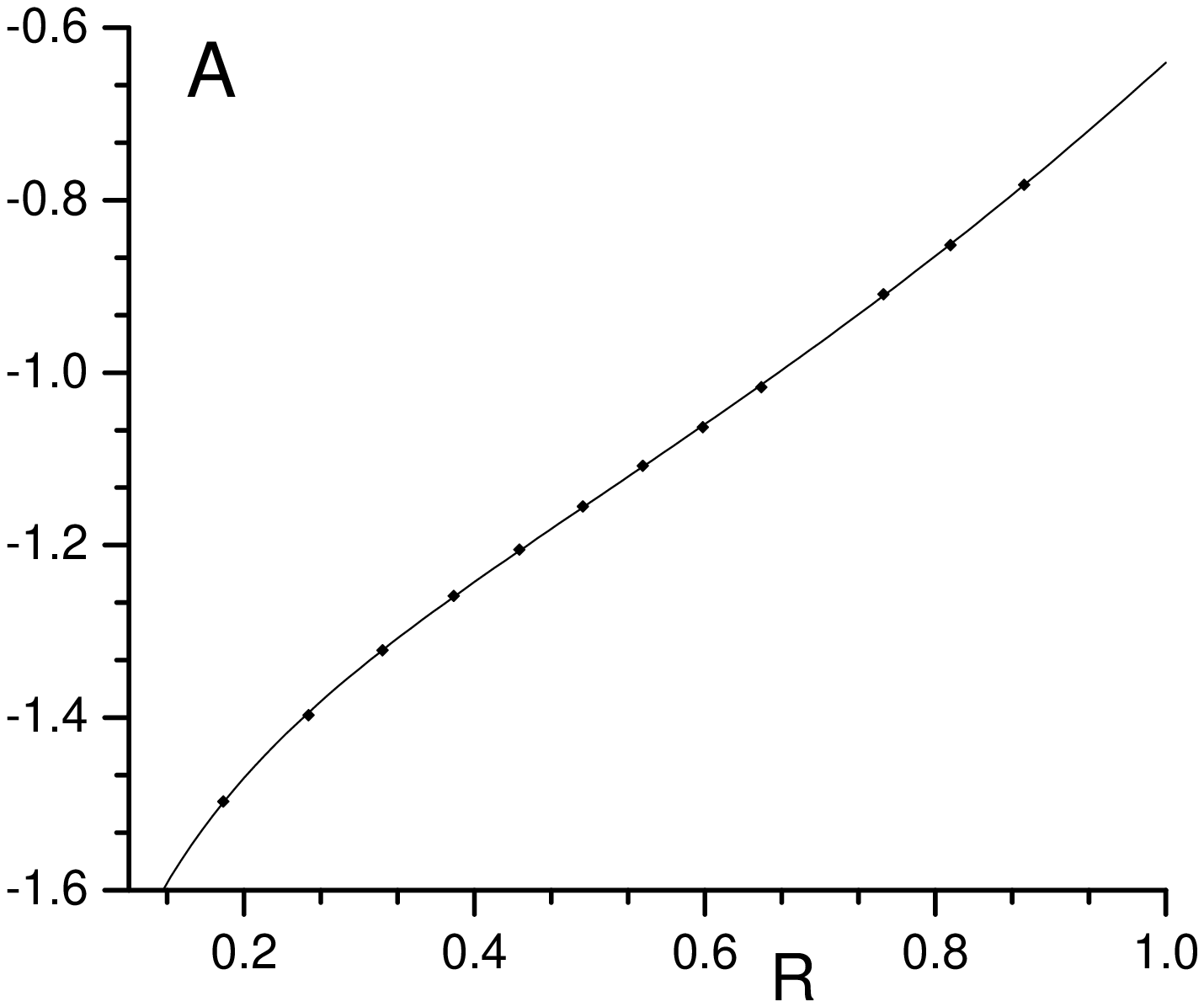}
\begin{minipage}{3in}
   \caption{Curves of proper area $A$ (in unit $3g_sN/2^{2/3}$) of
   string world-sheet vs. radius $R$ of Wilson loop at shorter
   distance. The diamonds denote numerical results, and solid line is
   the fit function, $A=-1.511-0.0.829R+0.0753\ln{R}$
   with $\chi^2=3\times 10^{-5}$.}
\end{minipage}}
\end{figure}
\end{itemize}

The main conclusions of this paper are included in fig.~3 and
fig.~4. In order to precisely describe behavior of
$A\sim\ln{<W({\cal C})}>$ with variation of radius $R$ of Wilson
loop, we fit numerical data with several different functions and
pick up the functions with the smallest $\chi^2$ to approach the
behaviors of Wilson loop.

The resulted functions for long distance (larger $R$) are listed
in Eq.~(\ref{19a})-(\ref{19c}), which fit the numerical data in
fig.~3. \\
 \alphaeqn
 \stepcounter{saveeqn}
\setcounter{equation}{0}
\renewcommand{\theequation}{\mbox{\arabic{saveeqn}-\alph{equation}}}
\parbox{1in}{\mbox{}}\hfill\parbox{5.5in}{
\parbox{3in}{\begin{eqnarray*}
  A&=&-1.5126+0.2941R+1.04R^2-2.162\ln{R}  \\
 A&=&-0.7835-1.0274R+1.1367R^2   \\
 A&=&-0.0464-1.717R+1.342R^2-0.014R^3
 \end{eqnarray*}}
 \parbox{2.5in}{
 \beq{19}\label{19a}&&\chi^2=0.0013 \\ \label{19b}&& \chi^2=0.0046 \\
 \label{19c}&& \chi^2=0.0017
 \eeq}}\stepcounter{saveeqn}
 \setcounter{equation}{0} \\

We can see that Eq.~(\ref{19a}) is the best approach on Wilson
loop at large distance. Therefore, it is obvious that Wilson loop
exhibits area law for large $R$. In Eq.~(\ref{19a}) we also meet
linear term and logarithmic term of $R$ which are unexpected.
These extra terms may be induced by the following reasons:
\begin{enumerate}
\item Due to error bar of numerical evaluation, the subtraction on
divergence in Eq.~(\ref{18}) is not precise.
\item Other error bar of numerical evaluation.
\item Due to lack of powerful computers, in our numerical
evaluation $R$ is not taken very large. Then resulted Wilson loop
does not lie in pure confinement phase. These extra may implies
some ``mixing-phase''.
\end{enumerate}
However, it is unambiguous that Wilson loop will be dominant by
$R^2$-term for very large $R$. Consequently $\net$ SYM is
confined at long distance.

Now let us pay attention to short distance (smaller $R$). The
numerical data and fit curve are shown in fig.~4, and fit
functions are listed as follows,
\\ \parbox{1in}{\mbox{}}\hfill\parbox{5.5in}{
\parbox{3in}{\begin{eqnarray*}
 A&=&-1.511+0.829R+0.0753\ln{R}   \\
 A&=&-1.682+1.128R-0.131R^2
 \end{eqnarray*}}
 \parbox{2.5in}{
 \beq{20}\label{20a}&&\chi^2=6.4\times 10^{-4} \\
 \label{20b}&& \chi^2=0.001
 \eeq}} \\
 \reseteqn
\renewcommand{\theequation}{\arabic{equation}}
It is unambiguous that Eq.~(\ref{20a}) is the best fit. Therefore, Wilson
loop of $\net$ SYM exhibits logarithmic behavior at short distance. It
indicates that $\net$ SYM is asymptotical freedom or approach to Coulomb
phase at short distance. In addition, we see that the constants in
Eq.~(\ref{19a}) and in Eq.~(\ref{20a}) are almost same. Then this constant
is harmless. It should be associated to renormalization or normalization
of Wilson loop. This also indicates fitting in Eq.~(\ref{19a}) and in
Eq.~(\ref{20a}) are consistent.

Another interesting issue is that a phase transition occurs when $R$
varies from small ones to large ones. This phase transition occurs between
the phase of asymptotical freedom and the phase of confinement of
non-abelian gauge theory. The transition point lies in region $0.5<R<2$ in
unit $3g_sN/2^{2/3}$. Consequently from Eq.~(\ref{8}) we can define
$\Lambda_{\rm SYM}=\ep^{2/3}/\alpha'$ as a scale associating to
confinement. Using the results listed in ref.\cite{KS00,HKO01,CH01}, we
have:
\begin{itemize}
\item The masses of glueball and Kaluza-Klein (KK) states scale as
 \beq{21} m_{\rm glueball}\;\sim\; m_{\rm KK}\;\sim \;
  \frac{\ep^{2/3}}{g_sN\alpha'}\;\sim\;\Lambda_{\rm SYM}/g_sN. \eeq
We usually expect $m_{\rm glueball}\;\sim\; m_{\rm
KK}>\Lambda_{\rm SYM}$ such that it requires smaller
$g_s\sim\gym^2$. In other words, the glueball is formed at near
perturbative region. Using NSVZ $\beta$ function\cite{NSVZ} for
$\net$ SYM, we have
 \beq{22} \frac{1}{\gym^2 N}=\frac{\frac{3}{16\pi^2}
  \ln{\mu^2/\Lambda_{\rm SYM}^2}+c_0}
  {1-2\frac{\ln{(\ln{\mu^2/\Lambda_{\rm SYM}^2}+c_0)}}
  {\ln{\mu^2/\Lambda_{\rm SYM}^2}+c_0}}, \eeq
where $c_0$ is a small constant regularizing divergence at
$\mu=\Lambda_{\rm SYM}$ and $\mu$ is the scale that glueball or
KK states are produced (roughly we can set $\mu\simeq m_{\rm
glueball}$). Since the singularity in KST solution is removed
through the blowing-up of the $S^3$ of $T^{1,1}$, more precisely,
we can include relevant coefficient to Eq.~(\ref{21}),
 \beq{23}m_{\rm KK}\simeq 3I^{-1/2}(s_0)\Lambda_{\rm SYM}/g_sN, \eeq
where $I(s_0)$ was defined Eq.~(\ref{5}) and $s_0$, which denotes
the lowest energy detected by certain Wilson loop, is zero-point
of $f(s)$. Considering radius/energy-scale relation for KST
background\cite{WH02}, in our nations $s\sim \mu/\mu_0$, where
the typical scale $\mu_0\simeq \Lambda_{\rm SYM}$ consequently it
corresponds to $R_0\simeq 1$ and $\bar{s}_0\simeq 0.8$. Then
defining $x=m_{\rm KK}/\Lambda_{\rm SYM}\simeq \mu/\Lambda_{\rm
SYM}$, we achieve the following equation
 \beq{24} x\simeq 3I^{-1/2}(\bar{s}_0/x)\frac{\frac{3}{16\pi^2}
 \ln{x^2}}{1-2\frac{\ln{\ln{x^2}}}{\ln{x^2}}}. \eeq
The solution of the above equation is $x\simeq 2.6$ such that we
obtain
 \beq{25} m_{\rm glueball}\;\simeq\; m_{\rm KK}\simeq 2.6
  \Lambda_{\rm SYM}. \eeq
\item The mass of the baryon scales as
 \beq{26} M_b\;\sim\; N\frac{\ep^{2/3}}{\alpha'}\;\sim\;
 N\Lambda_{\rm SYM}. \eeq
\item The gluino condensate of $\net$ SYM scales as
 \beq{27} <\lambda\lambda>\;\sim\; N\frac{\ep^2}{\alpha'^3}
        \;\sim\; N\Lambda_{\rm SYM}^3. \eeq
\end{itemize}

In QCD we knew that there is a so-called ``confinement scale'',
$\Lambda_{\rm QCD}\simeq 0.3GeV$. Extend to the above results to QCD, we
have
 \beq{28}
  m_{\rm glueball}&\simeq& 2.6\Lambda_{\rm QCD}\simeq 0.8GeV,
  \nonumber \\
  M_{\rm baryon}&\simeq& N_c \Lambda_{\rm QCD}\simeq 0.9GeV, \\
  <\bar{\psi}\psi>&\simeq& \Lambda_{\rm QCD}^3\simeq (0.3GeV)^3,
   \nonumber \eeq
where we use gluino condensate to mimic quark condensate in QCD, but we
ignore $N_c$ in Eq.~(\ref{27}) since those chiral quarks are fundamental
representation of gauge group. It is surprised that the above results
agree with phenomenology results of QCD well. It apparently means that
pure $\net$ SYM mimics some low energy behaviors of QCD even without
chiral multiplets.

\section{Summary and Comments}
\label{sec4}

Circular Wilson loop of $\net$ SYM from supergravity KST solution. The
Wilson loop exhibits area law at long distance and logarithmic law at
short distance. Therefore, $\net$ SYM lies in confinement phase at low
energy, and is asymptotical freedom or approach to Coulomb phase at high
energy. The phase transition occurs when energy varies. We also discuss
glueball mass, baryon mass and gluino condensate. It is shown that pure
$\net$ SYM mimics some low energy behaviors of QCD even without chiral
multiplets.

We would like to make a comment on MN background. In principle, the
extension the calculation of this paper to MN background is directly. The
Euler-Lagrangian equation in MN solution, however, does not possesses
initial condition like Eq.~(\ref{13}). Instead, it will be $f\sim$ fixed,
$f'\to\infty$ at boundary. Since another initial condition at zero-point
$r=r_0$ of $f(r)$ is $f\to 0$ fixed, $f'\to -\infty$, the numerical
evaluation on solution of Euler-Lagrangian is very difficult according to
criterion in Appendix. The study on this issue, however, is valuable
because it can help us to understand which phase of $\net$ SYM in UV is
dual to MN background. Some more discussions on MN background and its
noncommutative extension are in ref.\cite{MMP03}.

\appendix*
\section{Numerical Evaluations}

The main difficulty on numerical evaluations are from near-zero
point behavior of $f(r)$, i.e., $f'(r\to r_0)\to -\infty$. It
makes Euler-Lagrangian equation~(\ref{12}) be stiff for $r\to
r_0$. If we use Eq.~(\ref{13}), or more precisely, Eq.~(\ref{16})
with very small $r$ as initial conditions of numerical
evaluation. Due to inevitable error on initial condition in
numerical evaluation, this error will be enlarged rapidly for
$r\to r_0$. Consequently we can not obtain a convergent solution.
Fortunately, since we can analytically obtain solution of
Eq.~(\ref{12}) for $r\to r_0$, we can use Eq.~(\ref{17}) as
initial condition of numerical evaluation. In this case, the
error bar induced by numerical approximation on initial condition
is controllable, and it will decrease when $r$ is evaluated from
$r_0$ to zero. So that we obtain the convergent result.

In our evaluation, we adopt the forth order Runge-Kutta method. We
take initial condition at $r_0-r=10^{-6}$ and interval $2\times
10^{-9}$. According to Eq.~(\ref{17}), the error bar on initial
condition is about $10^{-6}$. When we let initial value vary
$0.1\%$, the resulted $f(0)=R$ and area only vary $0.0001\%$. It
indicates that the numerical evaluation is indeed convergent, and
subleading order on initial condition~(\ref{17}) can be ignored
consistently.

Another detail is to fix the value of cut-off $\lambda$ defined in
Eq.~(\ref{18}). Theoretically, function $f'(r)$ is single-valued
in region $0\leq r\leq r_0$, and increases from $-\infty$ to zero
when $r$ varies from $r_0$ to zero. When string world-sheet lies
across branes, i.e., in region $r<0$, $f'(r)$ is again
single-valued and increases from $-\infty$ to zero when $r$
varies from $-r_0$ to zero. It means that $f'(r=0)$ is maximal
value of $f'(r)$ if we extend space to $-r_0\leq r\leq r_0$.
Numerically, however, we can not expect to obtain the maximal
value of $f'(r)$ at $r=0$ exactly and maximal value of $f'(r)$ is
not precise zero. Then we can take maximal-value point $f'(r)$ as
a natural cut-off. In our evaluation, the value of the cut-off
$\lambda$ varies from $0.0015$ to $0.0033$ when $R=f(0)$ varies
from $0.1$ to $5.5$. It implies that we indeed consider the
second term in Eq~(\ref{18}) for smaller $R$ in order to achieve
high precision fit.

\end{document}